\documentclass[final,5p,times,twocolumn]{elsarticle}

\usepackage{amsmath}
\usepackage{amssymb}
\usepackage{graphicx}
\usepackage{color}
\usepackage[hypertex]{hyperref}
\hypersetup{colorlinks=true, linkcolor=purple, citecolor= purple, urlcolor=blue}

\usepackage[mathscr,scaled=1.15]{urwchancal}
\DeclareFontFamily{OT1}{pzc}{}
\DeclareFontShape{OT1}{pzc}{m}{it}%
{<-> s * [1.15] pzcmi7t}{}
\DeclareMathAlphabet{\mathpzc}{OT1}{pzc}{m}{it}

\biboptions{sort&compress}

\journal{Physics Letters B}

\begin{document}

\begin{frontmatter}

\title{Leading-twist parton distribution amplitudes of $S$-wave heavy-quarkonia}

\author[PKU1,PKU2]{Minghui Ding}
\author[PKU1,PKU2]{Fei Gao}
\author[NKU]{Lei Chang}
\author[PKU1,PKU2,PKU3]{Yu-{X}in Liu }
\ead{corresponding authors - yxliu@pku.edu.cn}
\author[ANL]{Craig D. Roberts}
\ead{cdroberts@anl.gov}

\address[PKU1]{Department of Physics and State Key Laboratory of Nuclear Physics and Technology, Peking University, Beijing 100871, China}
\address[PKU2]{Collaborative Innovation Center of Quantum Matter, Beijing 100871, China}
\address[NKU]{School of Physics, Nankai University, Tianjin 300071, China}
\address[PKU3]{Center for High Energy Physics, Peking University, Beijing 100871, China}
\address[ANL]{Physics Division, Argonne National Laboratory, Argonne, Illinois
60439, USA}

\date{16 November 2015}
%\date{19 October 2015}
%\date{7 October 2015}
%\date{21 May 2015}

\begin{abstract}
The leading-twist parton distribution amplitudes (PDAs) of ground-state $^1S_0$ and $^3S_1$ $c\bar c$- and $b\bar b$-quarkonia are calculated using a symmetry-preserving continuum treatment of the meson bound-state problem which unifies the properties of these heavy-quark systems with those of light-quark bound-states, including QCD's Goldstone modes.
Analysing the evolution of $^1S_0$ and $^3S_1$ PDAs with current-quark mass, $\hat m_q$, increasing away from the chiral limit, it is found that in all cases there is a value of $\hat m_q$ for which the PDA matches the asymptotic form appropriate to QCD's conformal limit and hence is insensitive to changes in renormalisation scale, $\zeta$.  This mass lies just above that associated with the $s$-quark.
At current-quark masses associated with heavy-quarkonia, on the other hand, the PDAs are piecewise convex-concave-convex.  They are much narrower than the asymptotic distribution on a large domain of $\zeta$; but nonetheless deviate noticeably from $\varphi_{Q\bar Q}(x) = \delta(x-1/2)$, which is the result in the static-quark limit.  There are also material differences between $^1S_0$ and $^3S_1$ PDAs, and between the PDAs for different vector-meson polarisations, which vanish slowly with increasing $\zeta$.
An analysis of moments of the root-mean-square relative-velocity, $\langle v^{2m}\rangle$, in $^1S_0$ and $^3S_1$  systems reveals that $\langle v^4\rangle$-contributions may be needed in order to obtain a reliable estimate of matrix elements using such an expansion, especially for processes involving heavy pseudoscalar quarkonia.
\end{abstract}


%%\pacs{
%%14.40.Pq,   % Heavy quarkonia
%%11.10.St,   %	Bound and unstable states; Bethe-Salpeter equations
%%12.38.Aw,	% General properties of QCD (dynamics, confinement, etc.)
%%12.38.Lg   % 	Other nonperturbative calculations
%%}

\end{frontmatter}

\noindent\textbf{1.$\;$Introduction}.
In studying hard exclusive processes within the Standard Model there are many instances in which one may appeal to factorisation theorems so that, at leading-order in a systematic expansion, the amplitude involved can be written as a convolution of a hard-scattering kernel, calculable in perturbation theory, and the so-called leading-twist PDA of the hadron involved, $\varphi(x)$, where $x$ is the light-front fraction of the hadron's total momentum carried by the struck parton.  Well known examples are formulae for the large momentum-transfer (asymptotic) behaviour of the electromagnetic charged-pion elastic and neutral-pion transition form factors \cite{Lepage:1979zb, Efremov:1979qk, Lepage:1980fj, Farrar:1979aw}.

For mesons, the PDA is a light-front projection of the system's Bethe-Salpeter wave-function onto the light-front.  It is therefore process independent and hence plays a crucial role in explaining and understanding a wide range of a given meson's properties and interactions.  The PDA is also essentially nonperturbative, \emph{i.e}.\ it cannot be calculated using perturbation theory.  The last two decades have witnessed significant progress toward the computation of realistic meson Bethe-Salpeter amplitudes \cite{Maris:2003vk, Chang:2011vu, Bashir:2012fs, Cloet:2013jya}; and, critically, the last two years have seen the development of novel techniques which enable the reliable calculation of meson PDAs from such Bethe-Salpeter amplitudes \cite{Chang:2013pqS, Cloet:2013tta}.  Predictions are now available for the (leading) twist-two PDAs of the $\pi$-, $K$-, $\rho$- and $\phi$-mesons \cite{Segovia:2013ecaS, Shi:2014uwaS, Gao:2014bcaS}, and for twist-three pion and kaon PDAs \cite{Chang:2013epa, Shi:2015esaS}.

Given that the last fifteen years have seen a dramatic expansion of interest in heavy-quark systems, owing to advances in both theoretical methods, and experimental activity and discoveries \cite{Eichten:2007qx, Brambilla:2010cs}, it is an opportune moment to use the continuum approach indicated above in order to compute the twist-two PDAs of $S$-wave heavy-quarkonia.  The theoretical interest is plain: one thereby arrives at a unified description and explanation of the leading-twist PDAs for almost all empirically accessible pseudoscalar and vector mesons.  This means, \emph{e.g}.\ that one simultaneously obtains an understanding of the structure of QCD's Goldstone modes and the $\eta_b$-meson, and can track the structural rearrangements which take place as a growth in current-quark mass drives an evolution between them.  There is also a phenomenological imperative: heavy-quarkonia PDAs appear in the analysis of numerous hard exclusive processes, \emph{e.g}.\ quarkonia production at high-energies \cite{Braguta:2006wr, Braguta:2007fh}; $J/\Psi+ \eta_c$ pair production in $e^+\, e^-$ annihilation \cite{Braguta:2007fh, Choi:2007ze}; $B_c \to \eta_c$ transitions \cite{Zhong:2014fma}; decays of heavy $S$-wave quarkonia into lighter vector mesons \cite{Braguta:2009xu}; deeply virtual quarkonia production, which can be used to probe the gluon distribution in the proton \cite{Koempel:2011rc}; and Higgs boson decays into quarkonia \cite{Bodwin:2014bpa}.

Notably, it is often supposed that such and kindred processes may be treated accurately using non-relativistic QCD (NRQCD) \cite{Bodwin:1994jh}, wherewith the leading-twist quarkonia PDAs are approximated as $\varphi_{Q\bar Q}(x,\zeta) = \delta(x-1/2)$, where $\zeta$ is the scale of the momentum transfer involved, and effects of nonzero $Q$-$\bar Q$ relative velocity, $\langle v^2\rangle>0$, are treated perturbatively.  However, whereas this might be true for processes that only involve $b$-quarks, corrections as large as a factor of two or more have been found when $c$-quarks are involved \cite{Ivanov:1998ms, Ivanov:2007cw, ElBennich:2010ha, ElBennich:2011py}.  A calculation of $\varphi_{Q\bar Q}(x,\zeta)$ in a framework that is capable of unifying this PDA with those of light-mesons can therefore also serve as valuable check on the fidelity of the NRQCD approximation.  The Dyson-Schwinger equations (DSEs) \cite{Maris:2003vk, Chang:2011vu, Bashir:2012fs, Cloet:2013jya} have this feature and we use that framework herein.

\smallskip

\noindent\textbf{2.$\;$Parton distribution amplitudes}.
In calculating the leading-twist PDAs of heavy quarkonia, we follow Refs.\,\cite{Chang:2013pqS, Gao:2014bcaS} and consider projections of their Bethe-Salpeter wave functions onto the light-front:
%{\allowdisplaybreaks
\begin{subequations}
\label{momPDAs}
\begin{align}
\label{eq:pda1}
 &f_{P}\, \varphi_{P}(x,\zeta)\notag\\
= & {\rm tr}_{CD} \, Z_2(\zeta,\Lambda) \!  \int^{\Lambda}_{dq}\delta(n\cdot q_{+}-x n\cdot P)\gamma_5\gamma\cdot n\chi_{P}(q;P)\,, \\
\label{eq:pda3}
&f_{V} n\cdot P\, \varphi_{V}^\|(x,\zeta) \notag\\
=  &m_{V}\,{\rm tr}_{CD} \, Z_2(\zeta,\Lambda) \!  \int^{\Lambda}_{dq}\delta(n\cdot q_{+}-x n\cdot P)n\cdot\gamma n_\lambda\chi_\lambda(q;P)\,,\\
\label{eq:pda2}
&f^\bot_{V} m_V^2 \, \varphi_{V}^\bot(x,\zeta) \notag\\
= &n\cdot P\, {\rm tr}_{CD}\, Z_T(\zeta,\Lambda) \! \int^{\Lambda}_{dq}\delta(n\cdot q_{+}-x n\cdot P)\sigma_{\mu\lambda}P_\mu\chi_{\lambda}(q;P)\,,
\end{align}
\end{subequations}%}
%\hspace*{-0.5\parindent}
where: $P$, $V$ denote, respectively, pseudoscalar ($^1S_0$) and vector ($^3S_1$) quarkonia; the trace is over color and spinor indices; $\int_{dq}^\Lambda $ is a Poincar\'e-invariant regularisation of the four-dimensional integral, with $\Lambda$ the ultraviolet regularisation mass-scale; $Z_{2,T}(\zeta,\Lambda)$ are, respectively, the renormalisation constants for the quark wave-function and the tensor vertex, which we compute in the chiral limit at the renormalisation scale $\zeta$; $n$ is a lightlike four-vector; $P$ is the meson's four-momentum, with $P^2=-m_{P, V}^2$, and $n\cdot P=-m_{P, V}$, with $m_{P, V}$ being the meson's mass.

In writing Eqs.\,\eqref{momPDAs} we have used the fact that there are only two independent vector-meson PDAs at leading-twist \cite{Ball:1998sk}: $\varphi_V^\|(x)$, $\varphi_V^\bot(x)$ describe, respectively, the light-front fraction of the meson's total momentum carried by the quark in a longitudinally or transversely polarised bound-state.  We have also adopted the convention $\int_0^1 dx\,\varphi(x)=1$, so that $f_{P}$, $f_{V}$, $f^\bot_{V}$ are decay constants.  The first two are measurable; but, whilst the tensor couplings $f^T_{V}$ are gauge- and Poincar\'e-invariant, they depend on the renormalisation scale: $f^T_{V}(\zeta)\to 0$ as $\zeta\to \infty$.  (Further details are available in Appendix~A of Ref.\,\cite{Gao:2014bcaS}.)

The Bethe-Salpeter wave functions in Eqs.\,\eqref{momPDAs} can be written
\begin{subequations}
\begin{align}
\chi_P(q;P) &= S(q_+)\Gamma_{5}(q;P)S(q_{-})\,,\\
\chi_\lambda(q;P) & = S(q_+)\Gamma_\lambda(q;P)S(q_{-})\,,
\end{align}
\end{subequations}
with $\Gamma$ the relevant meson's Bethe-Salpeter amplitude and $S$ the dressed propagator for the quark in that bound-state.  We have defined $q_+=q+\eta P$, $q_-=q-(1-\eta)P$, $\eta\in [0,1]$.  Owing to Poincar\'e invariance, no observable can legitimately depend on $\eta$, \emph{i.e}.\ the definition of the relative momentum.  On the other hand, the choice $\eta=1/2$ is computationally convenient when solving the Bethe-Salpeter equation that describes a  bound-system constituted from equal-mass valence partons.

It is appropriate to remark here that in order to produce gauge invariant results, Eqs.\,\eqref{momPDAs} should contain contributions that derive from a Wilson line, ${\cal W}[-x n/2,x n/2]$, drawn between the bound-state's valence constituents.  The Wilson line vanishes in light-cone gauge and hence does not contribute when this choice is employed.  On the other hand, light-cone gauge is seldom practicable in either model calculations or quantitative nonperturbative analyses in continuum QCD.  Herein, as is typical in nonperturbative DSE studies, we employ Landau gauge because, \emph{inter alia} \cite{Bashir:2008fk, Bashir:2009fv, Raya:2013ina}: it is a fixed point of the renormalisation group; that gauge for which sensitivity to model-dependent differences between \emph{Ans\"atze} for the fermion--gauge-boson vertex are least noticeable; and a covariant gauge, which is readily implemented in numerical simulations of lattice-regularised QCD.  It is therefore significant that ${\cal W}[- xz/2,xz/2]$ is not quantitatively important in the calculation of leading-twist PDAs \cite{Kopeliovich:2011rv}.

With realistic meson Bethe-Salpeter amplitudes in hand, it is straightforward to follow Refs.\,\cite{Chang:2013pqS, Gao:2014bcaS} and obtain PDAs for heavy quarkonium bound-states from Eqs.\,\eqref{momPDAs}.  The first step is to compute these moments:
%{\allowdisplaybreaks
\begin{subequations}
\label{momentsE}
\begin{align}
\nonumber
\langle x^m\rangle_P & = \int_0^1 dx \, x^m\, \varphi_P(x) \\
&= \frac{1}{f_P}{\rm tr}_{\rm CD}
Z_2 \! \int_{dq}^\Lambda \!\!
\frac{(n\cdot q_+)^m}{(n\cdot P)^{m+1}} \,\gamma_5\gamma\cdot n\, \chi_P(q;P)\,,\\
%
%\nonumber
%\lefteqn{
\langle x^m \rangle_\| % = \int_0^1 dx \, x^m\, \varphi_V^\|(x)}\\
&=
\frac{m_V}{f_V}{\rm tr}_{\rm CD} Z_2
\int_{dq}^\Lambda \frac{[n\cdot q_+]^m}{[n\cdot P]^{m+2}}\,
\gamma\cdot n \,n_\lambda \chi_\lambda(q;P)\,,\quad
\label{momentsParellel}\\
%
%\nonumber
%\lefteqn{
\langle x^m \rangle_\perp %= \int_0^1 dx \, x^m\, \varphi_V^\perp(x)}\\
&= \frac{1}{f_V^\perp  m_V^2 }  {\rm tr}_{\rm CD} Z_T
\int_{dq}^\Lambda
\frac{[n\cdot q_+]^m}{[n\cdot P]^{m}}\, \sigma_{\mu\lambda} P_\mu \chi_\lambda(q;P)\,.\quad
\end{align}
\end{subequations}%}
%\hspace*{-0.5\parindent}
In our Poincar\'e-covariant framework, arbitrarily many moments can be calculated, in principle and practice.

Having computed a sufficient number of the moments for a light-quark system, one could then reconstruct the associated PDA using the ``Gegenbauer-$\alpha$'' procedure introduced in Refs.\,\cite{Cloet:2013ttaS, Chang:2013pqS}, which is ideal for representing the broad, concave amplitudes that are characteristic of such systems.
%the renormalisation scale $\zeta=\zeta_2 = 2\,$GeV.
For heavy quarkonia, on the other hand, one expects the PDAs to be piecewise convex-concave-convex on $x\in [0,1]$, as is typical of finite-width representations of $\delta(x-1/2)$; and hence we proceed by assuming that
%% x = (1+\xi)/2 ... 6 x (1-x) = (3/2) (1-xi^2)
%% a^2 -> (2/3) a^2
%% 2 a^2 -> (4/3) a^2 => 1/(2 a^2) -> 3 / ( 4 a^2)
\begin{subequations}
\label{phiQQbar}
\begin{align}
%%\varphi_I(\xi &=2x-1,a)  = \mathpzc{N}_{\,a} \, \tfrac{3}{2} (1-\xi^2) \, {\rm e}^{ \tfrac{3}{2} a^2 [(1-\xi^2) - 1]}\,,\\
%
%%\mathpzc{N}_{\,a}^{-1} & = \left(\sqrt{6 \pi } \left(3 a^2-1\right) \text{erf}\left(\sqrt{\tfrac{3}{2}}
%%   a\right)+6 a\, {\rm e}^{-3 a^2/2} \right)/(12 a^3)\,,
%%%
\varphi_I(\xi &=2x-1,a)  = \mathpzc{N}_{\,a} \, \tfrac{3}{2} (1-\xi^2) \, {\rm e}^{ a^2 [(1-\xi^2) - 1]}\,,\\
\mathpzc{N}_{\,a}^{-1} & = \left(3 \left(2 a^2-1\right) \sqrt{\pi }\, \text{erf}\left(a \right)+6 a\, {\rm e}^{- a^2} \right)/(8 a^3)\,,
%%\frac{3 \sqrt{\pi } \left(2 b^2-1\right) \text{erf}(b)+6 e^{-b^2} b}{8 b^3}
\end{align}
\end{subequations}
can serve as an efficacious replacement for the expansion of heavy-quarkonia PDAs in terms of order-$\alpha$ Gegenbauer polynomials.  This function should be viewed as an informed assessment of the likely prior-distribution, in the sense of a Bayesian analysis of the reconstruction problem.

At this point, with $2 m_{\rm max}$ moments computed for a given heavy-quarkonium state using the appropriate formula in Eq.\,\eqref{momentsE}, one determines $a$ in Eq.\,\eqref{phiQQbar} by minimising
\begin{subequations}
\label{ErrorTest}
\begin{align}
\label{epsilonI}
\epsilon_I^2 & = \tfrac{1}{m_{\rm max}} \sum_{l=1,2,\ldots,m_{\rm max}}[\langle \xi^{2l} \rangle_I/\langle \xi^{2l} \rangle-1]^2\,,\\
\langle \xi^{2l} \rangle_I  & = \int_0^1 d\xi\,\xi^{2l} \varphi_I(\xi,a)\,.
\end{align}
\end{subequations}
(N.B.\, $\varphi_{Q\bar Q}(\xi)=\varphi_{Q\bar Q}(-\xi)$ so all odd-power $\xi$-moments vanish.)
%
%If $\epsilon_I\lesssim 1\,$\%, then we judge our ``prior'' to be valid and hence the result to be a realistic representation of the heavy-quarkonium system's PDA.  For current-quark masses such that Eq.\,\eqref{epsilonI} yields $\epsilon_I \gtrsim  5$\%, we forego this procedure and instead employ the ``Gegenbauer-$\alpha$'' method of Refs.\,\cite{Cloet:2013ttaS, Chang:2013pqS}.

\smallskip

\noindent\textbf{3.$\;$Heavy quarkonia Bethe-Salpeter amplitudes}.
To continue with our calculation of $S$-wave heavy-quarkonia valence-quark PDAs, the dressed-quark propagators and Bethe-Salpeter amplitudes associated with these bound states are needed.  We compute these quantities using gap and Bethe-Salpeter equation solutions obtained using the rainbow-ladder (RL) truncation of QCD's DSEs \cite{Maris:2003vk, Chang:2011vu, Bashir:2012fs, Cloet:2013jya} and the interaction introduced in Ref.\,\cite{Qin:2011dd}.

The RL truncation is the leading order in a systematic, symmetry-preserving procedure that enables a tractable formulation of the continuum bound-state problem \cite{Munczek:1994zz, Bender:1996bb}.  It is widely used in hadron physics and known to be accurate for light-quark ground-state vector- and isospin-nonzero-pseudoscalar-mesons \cite{Maris:2003vk, Chang:2011vu, Bashir:2012fs, Cloet:2013jya}, and properties of the nucleon and $\Delta$-baryon \cite{Eichmann:2011ej, Chen:2012qr, Segovia:2013rca, Segovia:2013ugaS}, because corrections in these channels largely cancel owing to parameter-free preservation of relevant Ward-Green-Takahashi identities (WGTIs).

The RL truncation has also been explored in connection with heavy-light mesons and heavy-quarkonia \cite{Bender:2002as, Bhagwat:2004hn, Nguyen:2010yh, Blank:2011ha, Rojas:2014aka, Hilger:2014nma}.  Those studies reveal that beyond-RL corrections to the dressed--quark-gluon vertex and hence the Bethe-Salpeter kernel are critical in heavy-light systems; and an interaction strength for the RL kernel fitted to pion properties alone is not optimal in the treatment of heavy quarkonia.  Both observations are readily understood; but we focus on the latter because it is relevant herein.

The interaction in Ref.\,\cite{Qin:2011dd} is deliberately consistent with that determined in studies of QCD's gauge sector, which indicate that the gluon propagator is a bounded, regular function of spacelike momenta, $q^2$, that achieves its maximum value on this domain at $q^2=0$ \cite{Bowman:2004jm, Boucaud:2011ugS, Ayala:2012pb, Aguilar:2012rz}, and the dressed-quark-gluon vertex does not possess any structure which can qualitatively alter these features \cite{Skullerud:2003qu, Bhagwat:2004kj}.  It also preserves the one-loop renormalisation group behaviour of QCD so that, \emph{e.g}.\ the quark mass-function is independent of the renormalisation point, and the infrared behaviour is determined by a single parameter, conventionally expressed as $ (\varsigma_G)^3 := D\omega$. Computations \cite{Maris:2002mt, Qin:2011ddS, Qin:2011xqS} show that observable properties of light-quark ground-state vector- and isospin-nonzero pseudoscalar-mesons are practically insensitive to variations of $\omega \in [0.4,0.6]\,$GeV, so long as
\begin{equation}
 (\varsigma_G)^3 := D\omega = {\rm constant}.
\label{Dwconstant}
\end{equation}
(The midpoint $\omega=0.5\,$GeV is usually employed in calculations.)  This feature also extends to numerous properties of the nucleon and $\Delta$-baryon \cite{Eichmann:2012zz}.  The value of $\varsigma_G$ is chosen so as to obtain the measured value of the pion's leptonic decay constant, $f_\pi$; and in RL truncation this requires $\varsigma_G^{\rm RL} =0.87\,$GeV.

Following Ref.\,\cite{Chang:2009zb}, however, it has become possible to employ far more sophisticated kernels for the gap and Bethe-Salpeter equations, which overcome the weaknesses of RL truncation in all channels studied thus far.  This new technique, too, is symmetry preserving; but it has an additional strength, \emph{i.e}.\ the capacity to express dynamical chiral symmetry breaking (DCSB) nonperturbatively in the integral equations connected with bound-states.  That is a crucial advance because DCSB is an important emergent phenomena within the Standard Model: it is the origin of more than 98\% of the visible mass in the Universe \cite{Brodsky:2015aiaS}.  Owing to this feature, the new scheme is described as the ``DCSB-improved'' or ``DB'' truncation.  It preserves successes of the RL truncation; but has also enabled elucidation of many novel nonperturbative features of QCD \cite{Chang:2010hb, Chang:2011ei, Chen:2012qrS, Chang:2013pqS}.

In a realistic DB truncation, $\varsigma_G^{\rm DB}=0.55\,$GeV; a value which coincides with that predicted by solutions of gauge-sector gap equations in QCD \cite{Binosi:2014aea}.  Since all dressing of the quark-gluon vertex vanishes in the heavy-quark limit, so that RL truncation must become valid, then the aforementioned agreement entails both that $\varsigma_G^{\rm DB}$ should be the infrared mass-scale appropriate for the RL analysis of truly heavy-heavy systems and provide more realistic results in such treatments of empirically accessible heavy-quarkonia.  Herein we use both  $\varsigma_G^{\rm RL}$ and $\varsigma_G^{\rm DB}$; and, as will become clear from those comparisons with experiment which are possible, the expectations described here are confirmed.

With the kernels of the gap and Bethe-Salpeter equations specified, one can employ standard algorithms and obtain numerical solutions for the dressed-quark propagators and Bethe-Salpeter amplitudes we require.  Those solutions yield the predictions for quarkonia static properties listed in Table~\ref{tablestatic}.  The current-quark masses were chosen in order to fit $m_{\eta_c}$, $m_{\eta_b}$, and correspond to the values of the running masses listed in the table; and the Euclidean constituent-quark mass \cite{Maris:1997tm} $M_Q^E=\{p\, |\, M_Q(p^2)=p, p>0\}$, where $M_Q(p^2)$ is the momentum-dependent dressed-quark mass of a flavour-$Q$ quark.

\begin{table}[t]
\caption{Heavy quarkonia static properties computed using RL truncation with
$\varsigma_G^{\rm RL}=0.87\,$GeV and current-quark masses $m_c(\zeta_2)=1.21 \,GeV$, $m_b(\zeta_2)=4.19\,GeV$
$\Rightarrow$ $M_c(0)=1.63$, $M_c^E=1.35 \,GeV$ and $M_b(0)=4.52$, $M_b^E=3.89\,GeV$;
and $\varsigma_G^{\rm DB}=0.55\,$GeV and current-quark masses $m_c(\zeta_2)=1.22 \,GeV$, $m_b(\zeta_2)=4.17\,GeV$
$\Rightarrow$ $M_c(\zeta=0)=1.42$, $M_c^E=1.32 \,GeV$ and $M_b(0)=4.49$, $M_b^E=3.93\,GeV$.
For comparison, a survey of numerous analyses that use various other methods yields \cite{Agashe:2014kda}:
$m_{c}(m_c)=1.275 \pm 0.025\,$GeV, $m_{b}(m_b)=4.18 \pm 0.03\,$GeV;
and from data \cite{Agashe:2014kda} on $^1S_0 \to \gamma\gamma$ and $^3S_1 \to e^+ e^-$ decays, one may infer (in GeV):
$f_{\eta_c} = 0.238(12)$,
$f_{J/\Psi} = 0.294(5)$,
$f_{\Upsilon}=0.506(3)$,
which the third row, bottom panel, lists in simplified form.
In subsequent rows we list selected decay-constant results from:
lattice QCD (lQCD) \cite{Davies:2010ip, McNeile:2012qf, Donald:2012ga, Colquhoun:2014ica} --
$f_{\eta_c}=0.279(17)$, $f_{\eta_b}=0.472(4)$, $f_{J/\Psi}=0.286(4)$, $f_\Upsilon=0.459(22)$;
%
%mock meson approach \cite{Hwang:1997ie};
%and sum rules (SR) \cite{Zhong:2014fma}.
%
an earlier RL DSE study (DSE$_{10}$) \cite{Nguyen:2010yh}; and a constituent-quark model (CQM) \cite{Segovia:2008zz, SegoviaPrivate2015}. %
(All quantities in GeV and $f^\bot$ values are quoted at a renormalisation scale $\zeta=2\,$GeV$=:\zeta_2$.)
\label{tablestatic}
}
%       m_etac m_etab m_JPsi m_Upsling
% RL m_etac m_etab m_JPsi m_Upsling
% DB
\begin{tabular*}%{lcccc}
{\hsize}
{
l|@{\extracolsep{0ptplus1fil}}
c|@{\extracolsep{0ptplus1fil}}
c|@{\extracolsep{0ptplus1fil}}
c|@{\extracolsep{0ptplus1fil}}
c@{\extracolsep{0ptplus1fil}}}\hline
      & $m_{\eta_c}$ & $m_{\eta_b}$  & $m_{J/\Psi}$ & $m_{\Upsilon}$ \\\hline
$\varsigma_G^{\rm RL}$ & $2.98$ & $9.39$ & $3.26$ & $9.52\,$ \\\hline
$\varsigma_G^{\rm DB}$ & $2.98$ & $9.39$ & $3.07$ & $9.46\,$ \\\hline
expt.\,\mbox{\cite{Agashe:2014kda}} & 2.98 & 9.39 & 3.10 & $9.46\,$ \\\hline
\end{tabular*}

\smallskip

%
%       f_etac f_etab f_JPsi f_JPsi^T f_Upsilon f_Upsilon^T
% RL f_etac f_etab f_JPsi f_JPsi^T f_Upsilon f_Upsilon^T
% DB
%%%Are the f^Ts quoted at 2GeV?
\begin{tabular*}%{lcccc}
{\hsize}
{
l|@{\extracolsep{0ptplus1fil}}
c|@{\extracolsep{0ptplus1fil}}
c|@{\extracolsep{0ptplus1fil}}
c|@{\extracolsep{0ptplus1fil}}
c|@{\extracolsep{0ptplus1fil}}
c|@{\extracolsep{0ptplus1fil}}
c@{\extracolsep{0ptplus1fil}}}\hline
      & $f_{\eta_c}$ & $f_{\eta_b}$  & $f_{J/\Psi}$ & $f_{J/\Psi}^\bot$ & $f_{\Upsilon}$ & $f_{\Upsilon}^\bot$ \\\hline
$\varsigma_G^{\rm RL}$
&\, 0.389 \,&\, 0.597 \,&\,0.410\, &\,0.337\,&\,0.552\,& \,0.489\, \\\hline
$\varsigma_G^{\rm DB}$
& \, 0.262 \,&\, 0.543 \,&\,0.255\, &\,0.213\,&\,0.471\,& \,0.421\,\\\hline
Expt. % \,\cite{Agashe:2014kda}
    &\, 0.238 \,&\,   \,& \, 0.294 \, &\, &\,0.506\, & \, \\\hline
lQCD % \cite{Davies:2010ip, McNeile:2012qf, Donald:2012ga, Colquhoun:2014ica}
&\, 0.279 \,&\, 0.472 \, & \, 0.286 \, & \, & \,0.459 \, & \, \\\hline
%
%%-old \cite{Chiu:2007km}&\, 0.309 \,&\, 0.566 \, & \,  & \, & \, & \, \\\hline
%\cite{Hwang:1997ie}&\, 0.296 \,&\, 0.498 \,&\, 0.287\, & \, & \,0.502\,& \,\\\hline
%SR &\, 0.320 \,&\, 0.573 \,& \,  &\, &\, & \, \\\hline
%
DSE$_{10}$
&\, 0.274 \,&\, 0.489 \, & \, 0.293 \, & \, & \,0.482 \, & \, \\\hline
CQM
&\, 0.841 \,&\, 0.728 \, & \, 0.346 \, & \, & \,0.469 \, & \, \\\hline
\end{tabular*}
\end{table}
%%Chiu:2007km = chiu2007beauty
%%Hwang:1997ie=hwang1997decay
%%Zhong:2014fma=zhong2014heavy ... 1/Sqrt[2] is correctly accounted for
%%%lQCD
%%% 1008.4018 f_etac=0.3947(24)/Sqrt[2]=0.279(17)
%%% 1207.0994 f_etab=0.667(6)/Sqrt[2]=0.472(4)
%%% 1208.2855 f_J/Psi=0.405(6)/Sqrt[2]=0.286(4) (see Eq.(8) for decay formula)
%%% 1408.5768 f_Y=0.649(31)/Sqrt[2]=0.459(22)
%%% g+g decays and eta decay constants 0804.2180 Eq.(6)

The leptonic decay constants in Table~\ref{tablestatic} computed using $\varsigma_G^{\rm DB}$ differ with experiment by a 10\% root-mean-square relative-error (rms-re) and by the same amount when compared with the lQCD values, which themselves differ from experiment by 11\%.  Measured this way, they are similar to the DSE results in Refs.\,\cite{Nguyen:2010yh,Blank:2011ha}.  On the other hand, the rms-re between the $\varsigma_G^{\rm RL}$ results and experiment is 43\%.
The CQM values listed in the last row emphasise the difficulty, understood by practitioners \cite{Segovia:2013sxa}, that such approaches encounter when attempting to simultaneously describe light- and heavy-quarkonia in the absence of a veracious expression of relevant WGTIs \cite{Maris:1997hd, Qin:2014vya}.  (\emph{N.B}.\ Our numerically determined Bethe-Salpeter wave-functions for heavy $^1S_0$ quarkonia satisfy Eq.\,(18) in Ref.\,\cite{Maris:1997hd}, a corollary of the axial-vector WGTI, with an accuracy of $98.9\pm 0.4$\%.)

It is interesting to note that in the limit of infinitely-heavy quarks, when all other mass-scales can be neglected, one has \cite{Chay:2000xn, Bhagwat:2006xi, Braguta:2009ej}:
$m_P = 2 M_Q$, where one may choose $M_Q=M_Q(0)$, \emph{i.e}.\ the value of the appropriate dressed-quark mass-function at the origin, since this function no longer runs;
and $2 m_Q(\zeta) f_V = m_V f_V^\bot(\zeta)$.  Deviations from these predictions are one crude measure of the impact of essentially dynamical effects in heavy-quarkonia.  Working from Table~\ref{tablestatic}, $\varsigma_G^{\rm RL}$ results differ from these expectations with a 7\% rms-re, whilst the $\varsigma_G^{\rm DB}$ rms-re is 4\%.

\smallskip

\begin{table}[t]
\caption{Moments of heavy-quarkonia twist-two PDAs, evaluated at a renormalisation scale $\zeta=\zeta_2$. The upper panel was obtained with $\varsigma_G^{\rm RL}=0.87\,$GeV and the lower panel with $\varsigma_G^{\rm DB}=0.55\,$GeV.
\label{Table:moments}
}
\begin{tabular*}%{lcccc}
{\hsize}
{
l|@{\extracolsep{0ptplus1fil}}
l|@{\extracolsep{0ptplus1fil}}
l|@{\extracolsep{0ptplus1fil}}
l|@{\extracolsep{0ptplus1fil}}
l@{\extracolsep{0ptplus1fil}}}\hline
$\varsigma_G^{\rm RL}$ & $\langle \xi^2 \rangle$ & $\langle \xi^4 \rangle$ & $\langle \xi^6 \rangle$ & $\langle \xi^8 \rangle$ \\\hline
$\eta_c$ & 0.11$\phantom{7}$ & 0.045 & 0.022 & 0.0062 \\\hline
$\eta_b$ & 0.069 & 0.015 & 0.0035 & 0.0012 \\\hline
$J/\Psi$
\; $\bot$ & 0.037 & 0.0046 & $6.4 \times 10^{-4}$ & $1.7 \times 10^{-4}$\\
$\phantom{J/\Psi}$
\; $\|$    & 0.0064 & $0.0012$ & $4.1 \times 10^{-6}$ & $1.6 \times 10^{-7}$\\\hline
$\Upsilon$
\;\;\;\; $\bot$ & 0.020 & $8.9 \times 10^{-4}$ & $5.1 \times 10^{-5}$ & $9.1 \times 10^{-6}$\\
$\phantom{\Upsilon}$
\;\;\;\; $\|$    & 0.0022 & $2.1 \times 10^{-5}$ & $2.6 \times 10^{-7}$ & $5.5 \times 10^{-9}$\\\hline
\end{tabular*}

\smallskip

\begin{tabular*}%{lcccc}
{\hsize}
{
l|@{\extracolsep{0ptplus1fil}}
l|@{\extracolsep{0ptplus1fil}}
l|@{\extracolsep{0ptplus1fil}}
l|@{\extracolsep{0ptplus1fil}}
l@{\extracolsep{0ptplus1fil}}}\hline
$\varsigma_G^{\rm DB}$ & $\langle \xi^2 \rangle$ & $\langle \xi^4 \rangle$ & $\langle \xi^6 \rangle$ & $\langle \xi^8 \rangle$ \\\hline
$\eta_c$ & 0.10 & 0.032 & 0.015 & 0.0059 \\\hline
$\eta_b$ & 0.070 & 0.015 & 0.0042 & 0.0013 \\\hline
$J/\Psi$
\; $\bot$ & 0.048 & 0.0063 & $0.0017$ & $4.7 \times 10^{-4}$\\
$\phantom{J/\Psi}$ \;
$\|$    & 0.039 & $0.0038$ & $7.3 \times 10^{-4}$ & $3.3 \times 10^{-4}$\\\hline
$\Upsilon$
\;\;\;\; $\bot$ & 0.024 & 0.0010 & $5.9 \times 10^{-5}$ & $1.7 \times 10^{-5}$\\
$\phantom{\Upsilon}$
\;\;\;\; $\|$    & 0.014 & $4.3 \times 10^{-4}$ & $4.4 \times 10^{-5}$ & $3.7 \times 10^{-6}$\\\hline
\end{tabular*}

\end{table}

\begin{table}[t]
\caption{Computed values of the width parameter ``$a$'' in Eq.\,\eqref{phiQQbar}, which characterises the twist-two PDA of each heavy-quarkonium system.
% $\varsigma_G^{\rm RL}=0.87\,$GeV - $\varsigma_G^{\rm DB}=0.55\,$GeV.
\label{aValues}
}
\begin{tabular*}%{lcccc}
{\hsize}
{
l|@{\extracolsep{0ptplus1fil}}
l|@{\extracolsep{0ptplus1fil}}
l|@{\extracolsep{0ptplus1fil}}
l|@{\extracolsep{0ptplus1fil}}
l|@{\extracolsep{0ptplus1fil}}
l|@{\extracolsep{0ptplus1fil}}
l@{\extracolsep{0ptplus1fil}}}\hline
 & $\eta_c$ & $\eta_b$ & $J/\Psi^\bot$ & $J/\Psi^\|$ & $\Upsilon^\bot$ & $\Upsilon^\|$ \\\hline
$a_{\rm RL}$ & 1.7 & 2.6 & 3.6 & 8.8 & 5.5 & 14 \\\hline
$a_{\rm DB}$ & 1.7 & 2.5 & 3.0 & 3.4 & 5.3 & $\phantom{1}$6.0 \\\hline
\end{tabular*}
\end{table}

\begin{figure}[t]
\centerline{\includegraphics[width=0.5\textwidth]{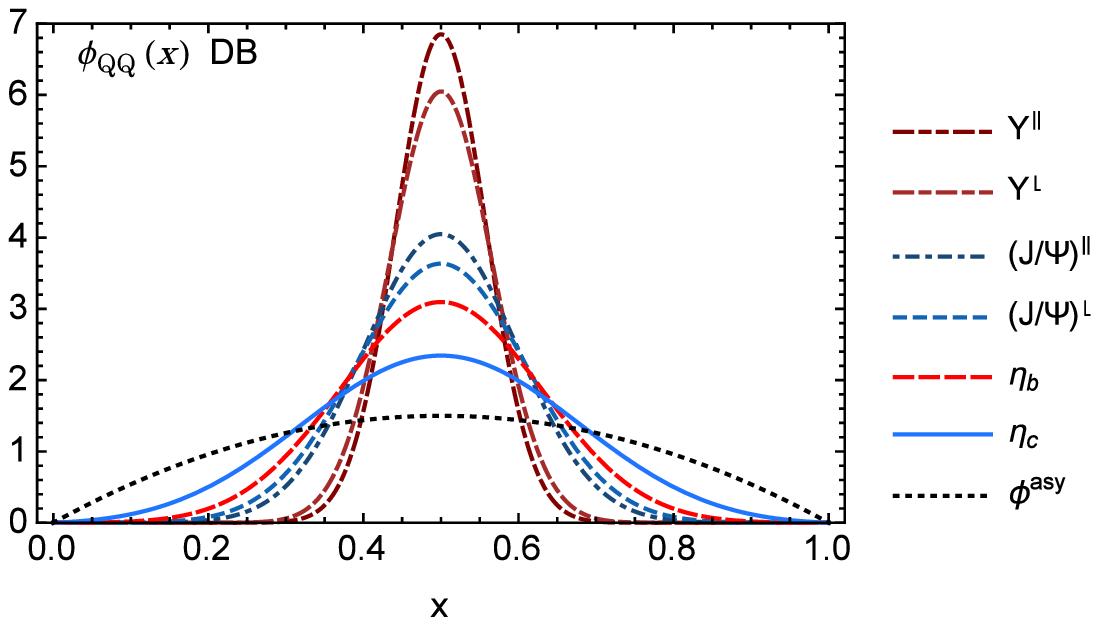}}
\centerline{\includegraphics[width=0.5\textwidth]{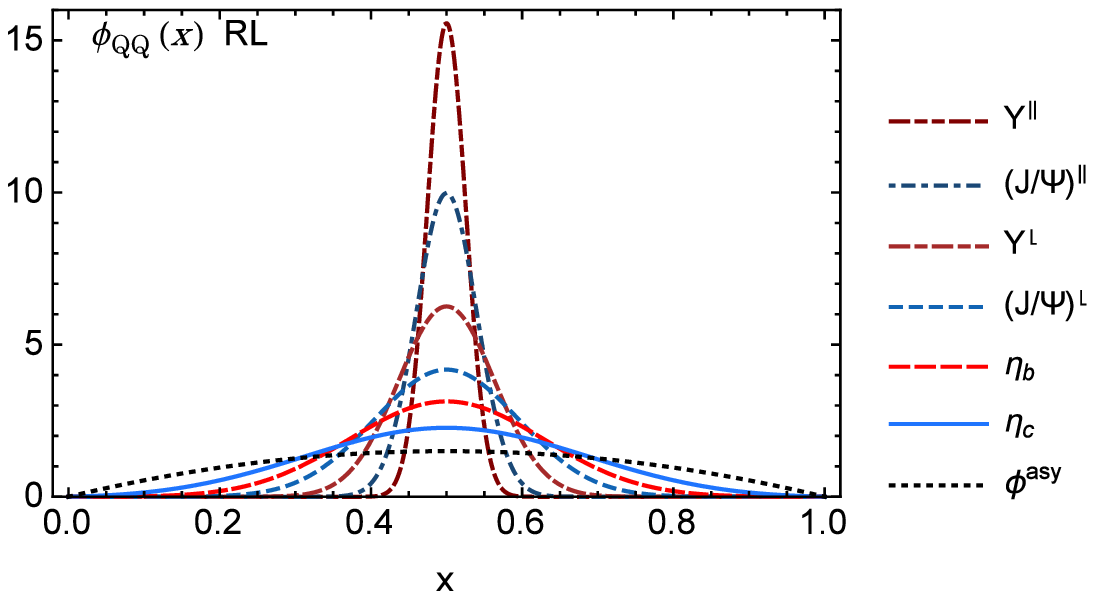}}
\caption{\label{fig:pda}  Twist-two heavy-quarkonia PDAs: upper panel, obtained using $\varsigma_G^{\rm DB}=0.55\,$GeV; and lower panel, obtained using $\varsigma_G^{\rm RL}=0.87\,$GeV.  The legend in each panel is ordered according to the maximum PDA peak-height; and $\varphi^{\rm asy}(x)=6 x(1-x)$, \emph{i.e}.\ the PDA associated with QCD's conformal limit \cite{Lepage:1979zb, Efremov:1979qk, Lepage:1980fj}.}
\end{figure}

\noindent\textbf{4.$\;$Twist-two PDAs}.
The procedure described in connection with Eqs.\,\eqref{momentsE}--\eqref{ErrorTest} can now be used to compute the  twist-two PDAs.   For systems composed of light-quarks, \emph{i.e}.\ those with current masses $\lesssim 0.1\,$GeV, which is roughly the $s$-quark mass, the $(n\cdot q_+)^m$ factor in Eqs.\,\eqref{momentsE} produces a highly-oscillatory integrand and thus reliable values for the moments cannot be obtained using a direct approach to computing the integrals.  In these cases, the procedure described in Ref.\,\cite{Chang:2013pqS}, based on generalised spectral representations of the light-quark propagators and bound-state amplitudes, is necessary and efficacious.  With increasing current-quark mass, however, owing to a damping influence from the large quark mass, this problem is shifted to progressively higher moments, which are also of diminishing magnitude and hence have little real impact, so that a ``brute-force'' approach is feasible for heavier quarkonia.

That is how we proceed with Eqs.\,\eqref{momentsE}--\eqref{ErrorTest} herein, \emph{viz}.\ direct integration using interpolations of numerical solutions for the propagators and Bethe-Salpeter amplitudes.  In order to eliminate dependence on the upper-bound of the momentum integration, which is a remnant of the oscillation problem just described, we introduced a factor $1/(1+k^2 r^2)^m$ for each moment $\langle \xi^{2m}\rangle$; computed the moment as a function of $r$; and extrapolated to $r=0$.  This procedure produced the values in Table~\ref{Table:moments}.

It is evident from Table~\ref{Table:moments} that in heavy-quarkonium systems one obtains a reasonable nonzero signal for moments $m\leq m_{\rm max}=4$.  Using these tabulated moments, Eq.\,\eqref{ErrorTest} yields twist-two PDAs of the form in Eq.\,\eqref{phiQQbar} with the width-parameters ``$a$'' in Table~\ref{aValues}.  In performing the least-squares fit we found $\epsilon_I^{\rm RL} = 20 \pm 14$\% and $\epsilon_I^{\rm DB} = 16\pm 9$\%.  These numbers can serve as an estimate of the errors in our moments and the associated reconstructions.  (\emph{N.B}.\ We examined a number of alternatives to Eqs.\,\eqref{phiQQbar}, including $\varphi_I(\xi) \propto (1-\xi^2)^\alpha \exp[a^2(1-\xi^2)-1]$, $\alpha \geq 1$, with no material rms-re improvement.)

The PDAs obtained with the widths in Table~\ref{aValues} are depicted in Fig.\,\ref{fig:pda}.  Consistent with the pattern that has already been established, we consider the results obtained using $\varsigma_G^{\rm DB}$, depicted in the upper panel, to be the more realistic; and, notably, peak-heights and widths in this case show a natural ordering:
\begin{equation}
\varphi_{\Upsilon^\|} <_N \varphi_{\Upsilon^\bot} <_N \varphi_{J/\Psi^\|} <_N \varphi_{J/\Psi^\bot}
<_N \varphi_{\eta_b} <_N \varphi_{\eta_c}  <_N \varphi^{\rm asy},
\end{equation}
where ``$<_N$'' means ``'narrower than''.

The PDAs obtained with $\varsigma_G^{\rm RL}$, drawn in the lower panel of Fig.\,\ref{fig:pda}, are anomalous in a number of ways, \emph{e.g}.\ the $\|$-PDAs are unnaturally sharply peaked and hence $\varphi_{J/\Psi^\|} <_N   \varphi_{\Upsilon^\bot}$.  Such behaviour can be traced to an over-concentration of interaction-strength in the far-infrared when one requires a good description of light-meson observables using RL truncation \cite{Chang:2013pqS, Cloet:2013tta, Segovia:2013ecaS, Shi:2014uwaS, Shi:2015esaS}.  This is corrected when DB kernels are employed \cite{Binosi:2014aea}; and that improvement is mimicked in RL-studies of heavy-quarkonia which broaden the interaction by increasing $\omega$ in Eq.\,\eqref{Dwconstant} \cite{Hilger:2014nma}.

%\noindent\textbf{5.$\;$Fixed point of the ERBL equation}.
%
Consideration of Fig.\,\ref{fig:pda} reveals a curious interplay between PDA evolution with current-quark mass and renormalisation scale.  Plainly, as $\Lambda_{\rm QCD}/\zeta\to 0$, where $\Lambda_{\rm QCD}$ is QCD's fixed renormalisation-induced scale, all twist-two meson PDAs must approach $\varphi^{\rm asy}(x)$.  On the other hand, for any fixed $\zeta$, $\varphi_{Q\bar Q}(x;\zeta)\to\delta(x-1/2)$ as $\Lambda_{\rm QCD}/m_Q(\zeta)\to 0$.  Given that each light-quarkonium PDA is a concave function of unit area which is broader than $\varphi^{\rm asy}(x)$ and ERBL evolution \cite{Lepage:1979zb, Efremov:1979qk, Lepage:1980fj} is smooth, uniform and area-preserving, then with increasing current-quark mass there must be a critical value, $m_q^c(\zeta)$, corresponding to a particular value of the renormalisation-group-invariant current-quark mass $\hat m_q^c$, at which $\varphi_{q\bar q}(x;\zeta) = \varphi^{\rm asy}(x)$.  So long as the current-quark mass remains  fixed by $\hat m_q^c$, then this light-quarkonim PDA cannot change under ERBL evolution, \emph{viz}.\ it is a fixed point.
This occurs at the following masses:
%The masses at which this occurs are:
\begin{equation}
\begin{array}{lccc}
                                                & \varphi_P & \varphi^\bot_V & \varphi^\|_V \\
%m^c_{\varsigma_G^{\rm RL}}  & 0.23 & 0.14 & 0.12 \\
m^c_{\varsigma_G^{\rm DB}}(\zeta_2)/\mbox{\rm GeV} & 0.15 & 0.13 & 0.12 \\
\end{array}\,.
\end{equation}
Whilst the actual values may change modestly in response to improvements of our framework, one may reliably conclude that the critical value typically lies just above the $s$-quark mass.

The moments in Table~\ref{Table:moments} can be used to compute the rms relative velocity of valence-constituents in the quarkonium system under consideration.  At leading-order of an expansion in this velocity \cite{Braguta:2006wr, Braguta:2009xu}: $\langle v^{2n}\rangle = (2n+1) \langle \xi^{2n}\rangle$.  Using $\varsigma_G^{\rm DB}$, we find:
\begin{equation}
\label{v2moments}
\begin{array}{lcccccc}
                            & \eta_c & \eta_b & J/\Psi^\bot & J/\Psi^\| & \Upsilon^\bot & \Upsilon^\| \\
\langle v^2 \rangle & 0.31 & 0.21 & 0.14 & 0.12 & 0.07 & 0.04\\
\langle v^4 \rangle & 0.16 & 0.07 & 0.03 & 0.02 & 0.01 & 0.00 \\
\langle v^6 \rangle & 0.11 & 0.03 & 0.01 & 0.01 & 0.00 & 0.00 \\
\end{array}\,.
\end{equation}
Evidently, the generalised Gremm-Kapustin relation \cite{Bodwin:2006dn}: $\langle v^{2n}\rangle \approx \langle v^2\rangle^n$, valid in nonrelativistic potential models, fails at every order for each system owing to the persistent $x=0,1$ endpoint tails of our computed PDAs.  Despite this, Eq.\,\eqref{v2moments} suggests that computation of physical observables using an expansion in rms relative velocity should converge reasonably quickly, but with $\langle v^4\rangle$-contributions that are sizeable for $^1S_0$ systems.

%% SR eta_c PDA ... 0611021 & J/\Psi PDA ... 0701234
%% Conclusion ... all PDAs are the same.
%% SR for eta_b Zhong:2014fma
Pointwise forms for the twist-two PDAs of $c\bar c$-quarkonia have been estimated using sum rules \cite{Braguta:2006wr, Braguta:2007fh, Zhong:2014fma} and a light-front CQM \cite{Choi:2007ze}.  Within their errors, those analyses report $\varphi_{\eta_c} \approx \varphi_{J/\Psi^\bot} \approx \varphi_{J/\Psi^\|}$.  Our framework does not suffer from this drawback.  There is little information on the PDAs of $b\bar b$-quarkonia; but a result for $\varphi_{\eta_b}$ can be inferred from Ref.\,\cite{Zhong:2014fma}.

\begin{figure}[t]
\centerline{\includegraphics[width=0.5\textwidth]{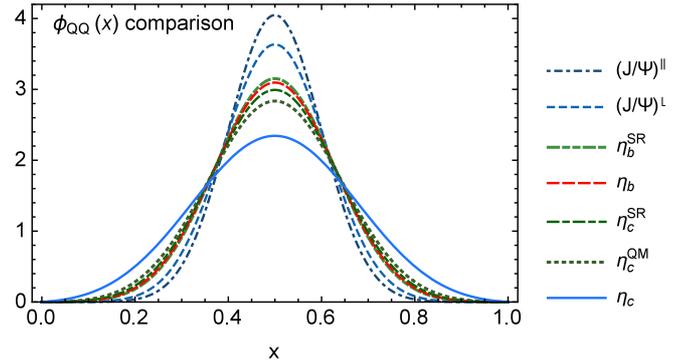}}
\caption{\label{fig:cfpda}  Comparison between our predictions for heavy-quarkonia PDAs and those reconstructed from moments computed using sum rules (SR) \cite{Braguta:2006wr, Braguta:2007fh, Zhong:2014fma} or a light-front CQM (QM) \cite{Choi:2007ze}.  Our predictions are the curves labelled $\eta_c$, $\eta_b$, $(J/\Psi)^\bot$, $(J/\Psi)^\|$.}
\end{figure}

We have taken the PDA moments reported in Refs.\,\cite{Braguta:2006wr, Braguta:2007fh, Zhong:2014fma, Choi:2007ze} and reconstructed pointwise forms for the associated heavy-quarkonia PDAs using the method described in connection with Eqs.\,\eqref{phiQQbar}, \eqref{ErrorTest} herein.  Figure~\ref{fig:cfpda} displays the resulting comparison.  The form of $\varphi_{\eta_b}$  determined from Ref.\,\cite{Zhong:2014fma} is almost identical to our prediction.  However, whilst the forms of $\varphi_{\eta_c}\approx \varphi_{J/\Psi^\bot} \approx \varphi_{J/\Psi^\|}$ reconstructed from Refs.\,\cite{Zhong:2014fma, Choi:2007ze} are quite similar to each other, they are markedly different from our predictions, especially insofar as we find material differences between these three systems.  (\emph{N.B}.\ Although Refs.\,\cite{Braguta:2006wr, Braguta:2007fh} used a function for $\varphi_I(\xi,a)$ in Eqs.\,\eqref{phiQQbar} that is somewhat different from ours, in comparing the results we found only immaterial pointwise differences.)

\smallskip

\noindent\textbf{6.$\;$Conclusion}.
We computed the leading-twist PDAs of $^1S_0$ and $^3S_1$ $c\bar c$- and $b\bar b$-quarkonia using a framework that has already provided explicit forms for PDAs of light-quark mesons, and hence arrived at a unified picture of systems ranging from QCD's Goldstone modes, whose properties are greatly influenced by dynamical chiral symmetry breaking, to those in which explicit chiral symmetry breaking is the dominant effect.

In this connection we examined the evolution of meson PDAs with current-quark mass, $\hat m_q$, and found that the broad, concave twist-two PDAs of light-quark systems change smoothly with increasing $\hat m_q$.  Consequently, there is always a value of $\hat m_q=:\hat m_q^c$ for which a given quarkonium PDA matches the asymptotic form, $\varphi^{\rm asy}(x)=6 x (1-x)$, and hence no longer evolves with renormalisation scale, $\zeta$.  This value of $\hat m_q$ lies just above that associated with the $s$-quark.  For $\hat m_q>\hat m_q^c$ the PDAs are piecewise convex-concave-convex; but naturally evolve to the fixed point $\varphi^{\rm asy}(x)$ as $\zeta$ is increased.

Heavy-quarkonia involve $\hat m_q \gg \hat m_q^c$, in which case the PDAs are piecewise convex-concave-convex and much narrower than $\varphi^{\rm asy}$ on a large domain of $\zeta$.  Nevertheless, for realistic values of $\hat m_q$ the PDAs deviate noticeably from $\varphi_{Q\bar Q}(x) = \delta(x-1/2)$, which is the limiting form in the case of static quarks.  There are also material differences between $^1S_0$ and $^3S_1$ PDAs and between the PDAs for different vector-meson polarisations, although these vanish slowly with increasing $\zeta$.
Considering these features, we computed moments of the rms relative-velocity, $\langle v^{2m}\rangle$, in $^1S_0$ and $^3S_1$  systems and found that in order to obtain accurate estimates for observables using an expansion in these terms it is likely that $\langle v^4\rangle$-corrections will require calculation, especially for processes involving $^1S_0$ systems.

\smallskip

\noindent\textbf{Acknowledgments}.
We are grateful for astute remarks by I.\,C.~Clo\"et, C.~Mezrag, S.-X.~Qin and J.~Segovia.
Work supported by:
National Natural Science Foundation of China under Contract Nos.\ 11435001 and 11175004;
the National Key Basic Research Program of China under Contract Nos.\ G2013CB834400 and 2015CB856900;
and the U.S.\ Department of Energy, Office of Science, Office of Nuclear Physics, under contract no.~DE-AC02-06CH11357.

%\bibliographystyle{../../zProc/z10/z10KITPC/h-physrev4}
%\bibliography{../../CollectedBiB}
%\bibliographystyle{../../zchanglei/zchanglei2/zPDA5/model1a-num-names}
%\bibliography{../../../CollectedBiB}

\end{document}